\documentclass[aps,prd,preprintnumbers,twocolumn]{revtex4}
\usepackage{graphicx}
\usepackage{amsmath}
\usepackage{bm}
\begin{document}

\voffset=.25in
\preprint{\vbox{ \hbox{ANL-HEP-CP-03-117}
                 \hbox{KIAS-P03091}
         }}
\title{ Exclusive two-charmonium vs. charmonium-glueball production at BELLE}
\thanks{Talk presented at 2nd International Conference on Flavor Physics 
(ICFP 2003), Korea Institute for Advanced Study, Seoul, Korea, 
October 6-11, 2003.}

\author{Jungil Lee}
\affiliation{High Energy Physics Division, Argonne National Laboratory,
Argonne, Illinois 60439, U.S.A.}
\affiliation{Korea Institute for Advanced Study, Seoul 130-722, Korea}

\date{\today}
\begin{abstract}
We review the current theoretical situation with regard to the 
anomalously large 
cross section for exclusive $J/\psi+\eta_c$ production
in $e^+e^-$ annihilation, as measured 
by BELLE Collaboration. 
\end{abstract}


\maketitle

\section{Introduction\label{intro}}
The $J/\psi$, a $J^{PC}=1^{--}$ charmonium state, has, for many years,
been a nice probe of hadron physics. Experimentally, it is easily
detectable through its decay into a lepton pair. Theoretically, it has
been a testing ground of perturbative quantum chromodynamics (QCD)
because it involves both long and short-distance dynamics. Because of
the separation between the long and short-distance scales involved in
quarkonium dynamics, one can make use of nonrelativistic QCD (NRQCD)
\cite{BBL} to describe the production and decay of heavy quarkonium in a
factorized form that is analogous with those in standard hard-scattering
QCD factorization theorems \cite{Collins:1987pm,Collins:gx,drell-yan}.
The factorized formula is a linear
combination of long-distance NRQCD matrix elements, which scale 
according to a known power of the 
velocity $v$ of the quark inside a quarkonium in the meson rest frame. 
The 
matrix elements are determined
by experimental measurements or, in some cases, through lattice 
calculations.
Once the universal long-distance matrix elements are determined, one
only needs to calculate the corresponding perturbative short-distance 
factors in order to predict the value of a physical observable.

As the first effective field theory for treating heavy quarkonium
physics, NRQCD succeeded in providing infrared-finite predictions of 
the $P-$wave quarkonium decay rate \cite{Bodwin:1992ye}
and presented a solution to the 
excessively large production rates for $J/\psi$ and $\psi'$ at large
transverse momenta at the Tevatron \cite{BF}.
These results were enough to rule out the color-singlet
model \cite{CSM}, which had
survived for decades. On the other hand, there are open questions still
to be resolved in NRQCD phenomenology. One is the polarization of the 
$J/\psi$ at large transverse momentum at the Tevatron: the CDF collaboration 
has not observed the large 
transverse polarization \cite{CDF-psi-pol} at large transverse momentum 
that was expected from theoretical predictions 
\cite{Cho-Wise,BKL-psi-pol,psi2}. However, the experimental data are 
not yet decisive. 
Experimental analyses with high-statistics, Run-II 
data are under way. On the theoretical side, 
large relativistic corrections have been found~\cite{large-v2}.
Furthermore, predictions have, so far, ignored spin-flip interactions. 
Lattice calculations to estimate the importance of these interactions 
are being carried out \cite{spin-flip}.

Recently, quarkonium physics has been faced with 
two particularly difficult problems,
both of which arise from BELLE measurements of $J/\psi$ production 
in $e^+e^-$ annihilation at $\sqrt{s}=10.6\,$GeV \cite{Abe:2002rb}.
One production rate for inclusive $J/\psi+c\bar{c}$
relative to $J/\psi+X$  \cite{Abe:2002rb}, which is much larger than 
predictions \cite{ee-prediction}.
The other is the cross section \cite{Abe:2002rb} for 
the exclusive process $e^+e^-\to J/\psi+\eta_c$, which is larger than 
predictions by
at least an order of magnitude~\cite{Braaten:2002fi,Liu:2002wq}. 
Both of them are the largest discrepancies that currently 
exist in the Standard Model.

In this proceeding, 
we review current theoretical status with regard to the exclusive problem.
We first consider the original discrepancy between 
the data  \cite{Abe:2002rb} and the theory \cite{Braaten:2002fi}
for exclusive $J/\psi+\eta_c$ production. Then we summarize two recently
proposed scenarios to reduce the discrepancy between the data and the theory.
One scenario is that the BELLE signal may contain double-$J/\psi$
events\cite{Bodwin:2002fk,Bodwin:2002kk}. 
The other is a conjecture that the signal may include exclusive
$J/\psi$ production associated with a glueball \cite{Brodsky:2003hv}.

\section{Exclusive \bm{$J/\psi+\eta_{\small{c}}$} production}
According to the NRQCD factorization formalism, exclusive processes of 
quarkonium production and decay, in which there is no other hadrons,
should be described by the color-singlet model, up to corrections 
that are higher order in $v$. As in the simplest examples, such as
electromagnetic annihilation decays and exclusive electromagnetic production
processes, $e^+ e^-$ annihilation into exactly two charmonia should be
accurately described by the color-singlet model, too.
Because of the mono-energetic nature of a two-body final state,
the absence of additional hadrons in the final state can be 
guaranteed experimentally.
For many charmonia $H$, the NRQCD matrix element can be determined 
from the electromagnetic annihilation decay rate of either $H$
or of another state that is related to $H$ by the heavy-quark spin symmetry.
Cross sections for exclusive two-charmonium production in 
$e^+e^-$ annihilation can, therefore, be predicted, 
up to corrections that are suppressed by powers of $v^2$,
without any unknown phenomenological factors. In contrast inclusive
quarkonium cross sections in hadroproduction involve several
matrix elements whose values have large uncertainties.  

Unfortunately, cross sections for exclusive two-charmonium production were too
small to be measured until the newly built high-luminosity $B$ factories 
collected large-statistics data. 
A naive estimate of the cross section
for $J/\psi + \eta_c$ in units of the cross section for $\mu^+ \mu^-$ is
\begin{eqnarray}
R[J/\psi + \eta_c]
\; \sim \; \alpha_s^2 \left( \frac{m_c v}{ E_{\rm beam}} \right)^6 .
\label{R-psieta:est}
\end{eqnarray}
The two powers of $\alpha_s$ are the minimum needed to produce a
$c \bar c + c \bar c$ final state.
There is a factor of $(m_c v)^3$ associated with the wave function
at the origin for each charmonium.
These factors in the numerator are compensated by factors of the
beam energy $E_{\rm beam}$ in the denominator, which yield a dimensionless 
ratio. As an example, consider $e^+ e^-$ annihilation with center-of-mass energy
$2 E_{\rm beam} = 10.6$ GeV.  If we set $v^2 \approx 0.3$,
$\alpha_s \approx 0.2$, and $m_c \approx 1.4$ GeV,
we obtain the naive estimate $R[J/\psi + \eta_c] \approx 4\times10^{-7}$.
This should be compared with the total ratio $R[{\rm hadrons}] \approx 3.6$
for all hadronic final states \cite{Ammar:1997sk}.

\begin{figure}
\includegraphics[height=8.5cm,angle=-90]{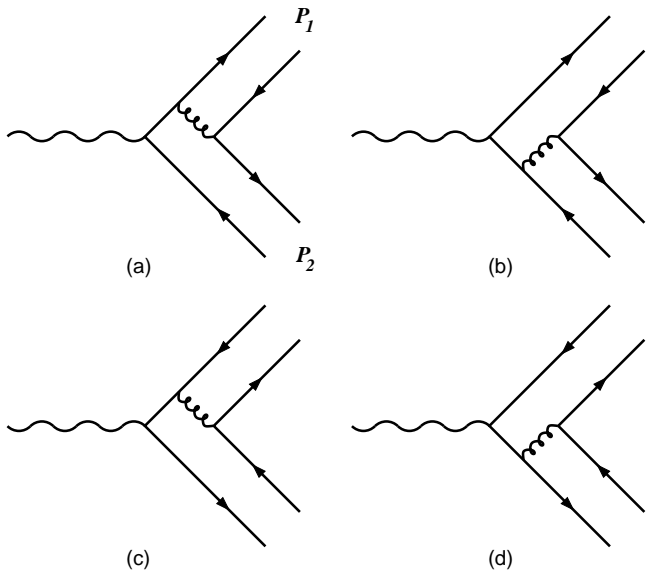}
\caption{\label{fig1}%
QCD diagrams that can contribute to the color-singlet process
$\gamma^* \to c \bar c_1 + c \bar c_1$.  }
\end{figure}
\begin{figure}
\includegraphics[height=8.5cm,angle=-90]{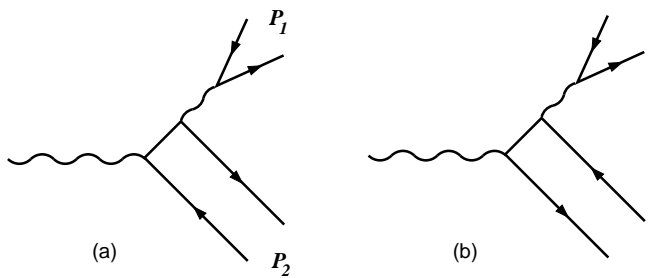}
\caption{\label{fig2}%
QED diagrams that contribute to the color-singlet process
$\gamma^* \to c \bar c_1(^3S_1) + c \bar c_1$.  }
\end{figure}

Because the two charmonia in the final state have opposite charge-conjugation
quantum numbers, it is assumed that they are produced by a single virtual 
photon.
The four QCD diagrams for the color-singlet process
$\gamma^* \to c \bar c_1 + c \bar c_1$ are shown in Fig.~\ref{fig1}.
We take the upper $c \bar c$ pair in Fig.~\ref{fig1}
to form a $C=-$ charmonium $H_1$ with momentum $P_1$ and the lower
$c \bar c$ pair to form a $C=+$ charmonium $H_2$ with momentum $P_2$.
There are also QED diagrams for
$\gamma^* \to c \bar c_1 + c \bar c_1$ that can be obtained from the
QCD diagrams in Fig.~\ref{fig1} by replacing the virtual gluons
by virtual photons, but they are suppressed by a factor of $\alpha/\alpha_s$.
However if one of the charmonia is a $1^{--}$ state, such as a $J/\psi$,
then there are the additional QED diagrams in Fig.~\ref{fig2}.
Although they are also  suppressed by a factor of $\alpha/\alpha_s$,
they are enhanced by a kinematic factor of $1/r^2$, where 
the variable $r$ is defined by
\begin{eqnarray}
r^2 &=& \frac{4 m_c^2 }{ E_{\rm beam}^2},
\label{r-def}
\end{eqnarray}
and therefore can be more important than one might expect.

The $\alpha^2 \alpha_s^2$ term in the cross section for 
$e^+ e^- \to J/\psi+\eta_c$
was calculated previously by Brodsky and Ji \cite{Brodsky:1985cr}.
They presented their result in the form of a graph of $R$ versus $1/r^2$,
but they did not give an analytic expression for the cross section.
Our predictions for the double-charmonium cross sections
without relativistic corrections
are given in Table~\ref{tab:sigma}.
The error bars are those associated with the uncertainty
in the NLO pole mass $m_c$ only.
\begin{table}
\caption{\label{tab:sigma}
        Cross sections in fb for $e^+ e^-$ annihilation into
        double-charmonium states $H_1+H_2$ at $E_{\rm beam} = 5.3$ GeV
        without relativistic corrections.
        The errors are only those from variations in the NLO pole mass
        $m_c = 1.4 \pm 0.2$ GeV. From Ref.~\cite{Braaten:2002fi}.}
\begin{ruledtabular}
\begin{tabular}{l|cc}
$H_2$ $\backslash$ $H_1$
& $J/\psi$ &$\psi(2S)$  \\
\hline
$\eta_c$
&2.31 $\pm$ 1.09&0.96 $\pm$ 0.45 \\
$\eta_c(2S)$
&0.96 $\pm$ 0.45&0.40 $\pm$ 0.19 \\
$\chi_{c0}(1P)$
&2.28 $\pm$ 1.03&0.95 $\pm$ 0.43 
\end{tabular}
\end{ruledtabular}
\end{table}
The relativistic corrections increase the  central values of the
cross sections by about
2.4 for $J/\psi+\eta_c$, by about 6 for $J/\psi+\eta_c(2S)$
and  $\psi(2S)+\eta_c$, and by about 13 for $\psi(2S)+\eta_c(2S)$.
Although the total correction factor for $J/\psi+\eta_c$
is significantly larger than 1, it is the product
of several modest correction factors that all go in the same
direction~\cite{Braaten:2002fi}.

The BELLE Collaboration has recently measured the cross section
for $J/\psi + \eta_c$ \cite{Abe:2002rb}.
The $J/\psi$ was detected
through its decays into $\mu^+\mu^-$ and $e^+e^-$,
which have a combined branching fraction of about 12\%.
The $\eta_c$ was observed as a peak in
the momentum spectrum of the $J/\psi$ corresponding to the 2-body
process $J/\psi + \eta_c$.
The measured cross section is
\begin{eqnarray}
\sigma[J/\psi+\eta_c] \times B[\ge 4]
= \left( 33^{+7}_{-6} \pm 9 \right) \; {\rm fb},
\label{BELLE}
\end{eqnarray}
where $B[\ge 4]$ is the branching fraction for the $\eta_c$ to decay
into at least 4 charged particles.  Since $B[\ge 4]<1$,
the right side of Eq.~(\ref{BELLE}) is a lower bound on the cross section
for $J/\psi + \eta_c$. Updated BELLE cross section is even 
larger than that given in Eq.~(\ref{BELLE})~\cite{BELLE-new}.

The lower bound provided by Eq.~(\ref{BELLE}) is about an order of magnitude
larger than the central value 2.3 fb of the calculated cross section
for $J/\psi + \eta_c$ in Table~\ref{tab:sigma}.
The largest theoretical errors are  QCD perturbative corrections,
which we estimate to give an uncertainty of roughly 60\%,
the value of $m_c$,
which we estimate to give an uncertainty of roughly 50\%,
and a relativistic correction that we have not been able to quantify
with confidence.

In Ref.~\cite{Braaten:2002fi}, we considered 
various two-charmonium processes in which the charmonia have opposite 
C-parity.
Complete helicity amplitudes for those processes are also presented
in Ref.~\cite{Braaten:2002fi}.
Liu, He, and Chao also calculated the $\alpha^2 \alpha_s^2$ terms
in the cross sections for
$e^+ e^-$ annihilation into $J/\psi + H$, $H$ = $\eta_c$, $\chi_{c0}$,
$\chi_{c1}$, and $\chi_{c2}$ \cite{Liu:2002wq}.
Their results are consistent with ours.

One might suspect that the discrepancy between theory and experiment
exists because of some problem in NRQCD factorization approach.
Substituting the input parameters used in above analysis into the
formula for the rate that is derived from light-front method in 
Ref.~\cite{Brodsky:1985cr}, we find the the result exactly agrees with that
from NRQCD \cite{Ji}. This implies that the discrepancy is not from
NRQCD, but from pQCD factorization itself.
\section{Exclusive double-\bm{$J/\psi$} production}
Having no conclusive idea for filling the huge gap between the data and theory
with perturbative corrections, we began to question whether the BELLE signal 
consists entirely of $J/\psi+\eta_c$ events.
As a possibly missing contribution,
Bodwin, Braaten, and I considered $e^+ e^-$ annihilation
into double-$J/\psi$  states that have the same charge-conjugation 
parity (C-parity)~\cite{Bodwin:2002fk,Bodwin:2002kk}. 
The strongest motivation was the observation that the width of the
$\eta_c$ signal measured by the BELLE Collaboration is similar to the mass 
difference between $J/\psi$ and $\eta_c$.
In the BELLE fit to the $J/\psi$
momentum distribution, the full width at half maximum of the $\eta_c$
peak is about 0.11 GeV. Since the mass difference between the $J/\psi$
and $\eta_c$ is about 0.12 GeV, there are probably $J/\psi + J/\psi$
events that contribute to the $J/\psi + \eta_c$ signal that is observed
by BELLE.

\begin{figure}
\includegraphics[width=6cm,angle=-90]{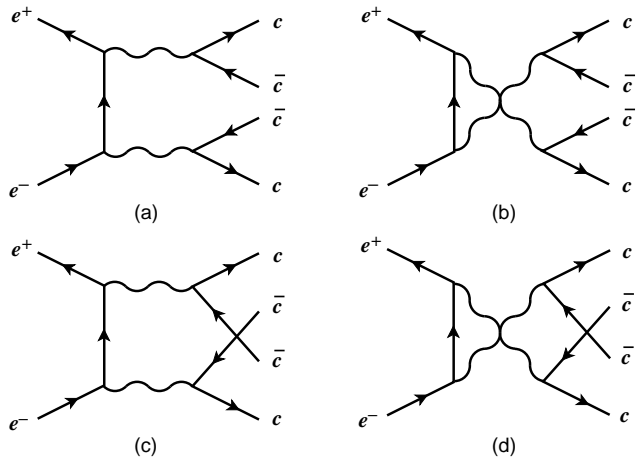}
\caption{\label{double2-f1}%
QED diagrams for the process $e^+ e^- \to \gamma^* \gamma^* \to c \bar
c_1 + c \bar c_1$. The upper and lower $c\bar{c}$ pairs evolve into
$H_1$ and $H_2$, respectively.}
\end{figure}

This process proceeds, at leading order in the QCD coupling $\alpha_s$,
through QED diagrams, shown in Fig.~\ref{double2-f1}, that contain two
virtual photons. One might expect that these cross sections would be
much smaller than those for charmonia with opposite C-parity because
they are suppressed by a factor of $\alpha^2/\alpha_s^2$. However, if
both charmonia have quantum numbers $J^{PC} = 1^{--}$, then there is a
contribution to the cross section in which each photon fragments into a
charmonium~\cite{Bodwin:2002fk,Bodwin:2002kk}.
The fragmentation contribution is enhanced by powers of 
$1/r$~\cite{Bodwin:2002fk,Bodwin:2002kk}. 
A similar, but less dramatic, example is the non-negligible QED contribution 
shown in Fig.~\ref{fig2} for $e^+e^-\to\gamma^*\to J/\psi+\eta_c$
process. This enhancement can compensate for the suppression factor that
is associated with the coupling constants.

The photon-fragmentation
contributions shown in Figs.~\ref{double2-f1}(a) and \ref{double2-f1}(b) 
are enhanced because the virtual-photon propagators are of
order $1/m_c^2$ instead of order $1/E_{\rm beam}^2$. In the amplitude,
there are also two numerator factors of $m_c$ instead of $E_{\rm beam}$,
which arise from the $c\bar c$ electromagnetic currents. Hence, the net
enhancement of the squared amplitude is $1/r^4$, and the
contributions to $R$ can be nonzero in the limit $r\to 0$. As $r \to
0$ with fixed scattering angle $\theta$, the photon-fragmentation
contributions to the cross section factor into the cross section for
$e^+ e^- \to \gamma \gamma$ with photon scattering angle $\theta$ and
the fragmentation probabilities for $\gamma \to H_1$ and $\gamma \to
H_2$. These fragmentation probabilities are nonzero at order $\alpha$
only for $J^{PC} = 1^{--}$ states with helicities satisfying $\lambda_1
= -\lambda_2 = \pm 1$. The contribution to the ratio $R$ for $J/\psi +
J/\psi$ has the behavior
\begin{eqnarray}
R[J/\psi(\lambda_1) + J/\psi(\lambda_2)]
\; \sim \; \alpha^2 (v^2)^{3},
\label{Rest:psi+psi}
\end{eqnarray}
where $\lambda_1 = -\lambda_2 = \pm 1$.
This asymptotic behavior can be compared with that for $J/\psi+\eta_c$:
\begin{eqnarray}
R[J/\psi + \eta_c]
\; \sim \; \alpha_s^2 (v^2)^{3} (r^2)^3,
\label{Rest:psi+eta}
\end{eqnarray}
which holds generally for $S$-wave final states with opposite C-parity. 
In this case, an additional factor $r^2$ arises from helicity suppression.
The ratio $R$ in Eq.~(\ref{Rest:psi+psi}) is suppressed relative to
Eq.~(\ref{Rest:psi+eta}) by a factor of $(\alpha/\alpha_s)^2 \approx
10^{-3}$, but the enhancement factor that scales as $r^{-6}$ makes the
cross sections comparable in magnitude at the energy of a $B$ factory.
In addition, the differential cross section has a sharp peak in
both forward and backward regions owing to the subprocess
$e^+e^-\to \gamma \gamma$, which results in another enhancement factor
$\ln(8/r^4)$ in the integrated cross section.

\begin{figure}
\includegraphics[height=6.7cm]{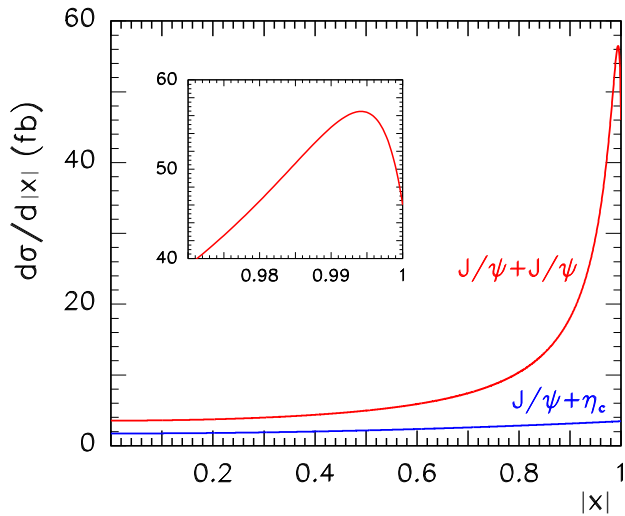}
\caption{\label{double2-f2}
Differential cross sections $d\sigma/d|x|$ for $e^+ e^-$ annihilation at
$E_{\rm beam}=5.3$~GeV into $J/\psi+J/\psi$ and $J/\psi+\eta_c$.
There are large errors associated with perturbative and
relativistic corrections, as discussed in the text. 
From Ref.~\cite{Bodwin:2002fk,Bodwin:2002kk}.}
\end{figure}

The differential cross section for exclusive double-$J/\psi$ production
as a function of $x=\cos\theta$ is shown
in Fig.~\ref{double2-f2}.
Our predictions for double-charmonium cross sections are given in
Table~\ref{tab:sigma--} for $C=-1$ states.
The error bars are those associated with the
uncertainty in the pole mass $m_c$ only.  
The cross sections for the $1^{--}$ states are
dominated by the photon-fragmentation diagrams in Figs.~\ref{double2-f1}(a)
and \ref{double2-f1}(b). For $m_c=1.4$~GeV, they contribute 87.5\% of the
$J/\psi+J/\psi$ cross section. The nonfragmentation diagrams in
Figs.~\ref{fig1}(c) and \ref{fig1}(d) contribute 0.7\%, while the
interference term contributes 11.8\%.

\begin{table}[t]
\caption{\label{tab:sigma--}%
Cross sections in fb for $e^+ e^-$ annihilation at $E_{\rm beam}=5.3$~GeV
into double-charmonium states $H_1+H_2$ with $C =-1$. The errors are
only those from variations in the pole mass $m_c = 1.4 \pm 0.2$~GeV.
There are additional large errors associated with perturbative and
relativistic corrections, as discussed in the text.
From Ref.~\cite{Bodwin:2002fk,Bodwin:2002kk}.}
\begin{ruledtabular}
\begin{tabular}{l|cc}
$H_2$ $\backslash$ $H_1$
&   $J/\psi$          &   $\psi(2S)$        \\
\hline
$J/\psi$
& 8.70   $\pm$ 2.94   &  7.22  $\pm$ 2.44   \\
$\psi(2S)$
&                     &  1.50  $\pm$ 0.51   
\end{tabular}
\end{ruledtabular}
\end{table}

The perturbative corrections to the cross section for $J/\psi + J/\psi$
have not yet been calculated. However the perturbative corrections to
the dominant photon-fragmentation diagrams in Figs.~\ref{double2-f1}(a) and
\ref{double2-f1}(b) are closely related to the perturbative correction to the
electromagnetic annihilation decay rate for $J/\psi \to e^+ e^-$, which
gives a multiplicative factor
\begin{eqnarray}
\left( 1 - \frac{8}{3} \, \frac{\alpha_s}{ \pi} \right)^2.
\label{gam-psi}
\end{eqnarray}
The perturbative correction to the photon-fragmentation terms in the
cross section for $J/\psi + J/\psi$ is just the square of the expression
(\ref{gam-psi}). If we choose the QCD coupling constant to be $\alpha_s%
= 0.25$, which corresponds to a renormalization scale $2 m_c$, then the
perturbative correction yields a multiplicative factor $(0.79)^4=0.39$.
The same perturbative correction factor applies to the cross sections
for $J/\psi + \psi(2S)$ and $\psi(2S) + \psi(2S)$. This perturbative
correction factor applies only to the leading contributions to the cross
sections in the limit $r \to 0$. However, since these contributions are
dominant, we conclude that the perturbative corrections are likely to
decrease the cross sections by about a factor of 3.

The relativistic corrections to the $J/\psi+J/\psi$ cross section are
significantly smaller than and have the opposite sign from the
relativistic corrections to the $J/\psi+\eta_c$ cross section, which are
given in Ref.~\cite{Braaten:2002fi}. The relativistic correction to
the fragmentation process is 0.78.
For $m_c = 1.4$ GeV, the
relativistic corrections to the $J/\psi+\eta_c$ cross section are
estimated to increase the cross section by about a factor 5.5. The large
difference in the relativistic corrections suggests that there may be
large relativistic corrections not only to the absolute cross sections
for double-charmonium production, but also to the ratios of those cross
sections. 

If we take into account both perturbative and relativistic corrections, 
then the predicted cross section for $J/\psi+J/\psi$ at the $B$ factories is 
of the same order of magnitude as that for $J/\psi+\eta_c$.
After this proposal was made, the BELLE Collaboration looked for the 
predicted double-$J/\psi$ events. 
Unfortunately, they did not find such events \cite{Abe:2003ja}.
The non-observation of double-$J/\psi$ 
events strongly suggests there is something wrong in our understanding of 
factorization in quarkonium production or of the relevant production 
mechanisms. If the factorization formalism is 
valid and we have accounted for the dominant production mechanisms,
then there is no reason why $J/\psi+\eta_c$ events should be seen while
double-$J/\psi$ events are not. Note that the two production processes depend 
on essentially the same non-perturbative factor.
At the moment, we know the relativistic correction seems to have an
important role in increasing the cross sections for  
$J/\psi+\eta_c$~\cite{Braaten:2002fi}.
The corrections to the two-charmonium processes in next-to-leading order 
in the strong coupling constant are still unknown. 
The NLO result may help us with pinning down the origin of the problem.
\section{Exclusive \bm{$J/\psi$}-glueball production}
The cross sections for $J/\psi+\eta_c$, $\chi_{c0}$, and $\eta_c(2S)$
recently measured by the BELLE Collaboration are
not well understood within pQCD or within NRQCD factorization.
If there is nothing wrong in our factorization formula, then
the BELLE signal must include something else that we have not considered.
The non-observation of double-$J/\psi$ events strongly suggests there is
something other than $\eta_c$ in the signal. This follows from the fact 
that the non-observation of double-$J/\psi$ events in the current BELLE data is
still consistent with the theoretical predictions for the absolute 
double-$J/\psi$ cross section. On the basis of this observation, 
Brodsky, Goldhaber, and I argued that the signal may include
events in which there is exclusive $J/\psi$ production associated with a 
$J^{PC}=J^{++}$ glueball $\mathcal{G}_J$ with $J=0,~2$~\cite{Brodsky:2003hv}.

Bound states of gluons provide an explicit signature of the
non-Abelian interactions of quantum chromodynamics.  In fact, in a
model universe without quarks, the hadronic spectrum of QCD would
consist solely of color-singlet glueball states. 
According to a recent  lattice
calculation by Morningstar and Peardon~\cite{Morningstar:1999rf},
the ground-state masses
for the $J^{PC}=0^{++}$ and $2^{++}$  glueballs $\mathcal{G}_J$ are
$1.73$ and $2.40$~GeV, respectively. There are many excited 
states in the mass range of charmonium spectrum~\cite{Morningstar:1999rf}.

The cross section for 
$e^+ e^- \to J/\psi +\mathcal{G}_J$ might not be suppressed 
compared with cross sections for exclusive quarkonium pairs such as 
$\gamma^* \to J/\psi +\eta_c$ that arise
from the subprocess $\gamma^* \to (c\bar c)+ (c \bar c)$. That is because
the  subprocesses
$\gamma^* \to (c\bar c) +(c \bar c)$  and
$\gamma^* \to (c \bar c) +(g g)$ are of the same nominal
order in pQCD. Thus, it is possible that
some portion of the signal observed by BELLE in $e^+ e^- \to J/\psi +X$
may actually be due to the production of 
$J/\psi +\mathcal{G}_J$ pairs.

The amplitude for $e^+ e^- \to J/\psi +\mathcal{G}_J$
at leading twist can be expressed as a
factorized product of the perturbative hard-scattering amplitude
$T_H(\gamma^* \to Q \bar Q +g g)$ convoluted with
the nonperturbative distribution amplitudes for the heavy quarkonium and
glueball states.
A bound on the normalization of the distribution amplitude for the 
glueball state can be extracted from a  resonance search by CUSB in
$\Upsilon\to \gamma +X$~\cite{CUSB-gamma} following the method used in
Ref.~\cite{Berger:2002bz}.

Our predictions for the upper limits to the exclusive
charmonium-glueball production cross sections are given in Table
\ref{tab:I0}. We find that the upper limit to the cross section
$\sigma_{J/\psi \mathcal{G}_0}$ is comparable to the NRQCD prediction of
the cross sections for $e^+e^-\to J/\psi+H$  for $H=\eta_c$ and
$\chi_{c0}$, and larger by factor 2 than that for $H=\eta_{c}(2S)$,
suggesting the possibility that a significant fraction of the
anomalously large cross section measured by BELLE may be due to 
glueballs in association with $J/\psi$ production. 

\begin{figure}
\includegraphics[height=4.5cm]{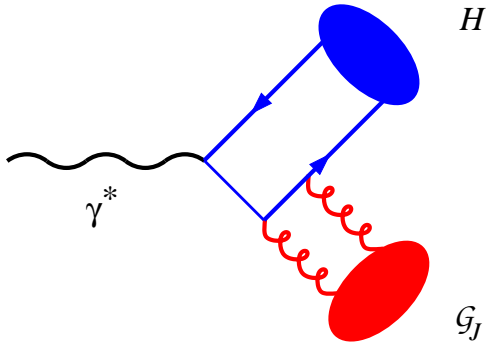}
\caption{\label{glue}%
Feynman diagram for
$\gamma^* \to H + \mathcal{G}_J$.  }
\end{figure}

\begin{table}[t]
\caption{\label{tab:I0}%
Upper limits to
the 
cross section $\sigma_{J/\psi \mathcal{G}_0}$
and the ratio
$\sigma_{J/\psi \mathcal{G}_0}/\sigma_{J/\psi H}$ at $\sqrt{s}=10.6$~GeV,
assuming that $M_{\mathcal{G}_0}=M_h$, 
where $H=$ $\eta_c$, $\chi_{c0}$, and $\eta_c(2S)$.
The limits are determined by
the $\Upsilon\to\gamma X$ search of the CUSB
Collaboration~\cite{CUSB-gamma}.
From Ref.~\cite{Brodsky:2003hv}.
}
\begin{ruledtabular}
\begin{tabular}{l|ccc}
$M_{\mathcal{G}_0}=M_h$ & $h=\eta_c$ & $\chi_{c0}$ & $\eta_c(2S)$
\\
\hline $\sigma^{\textrm{max}}_{J/\psi \mathcal{G}_0}$
& 1.4 fb  & 1.5 fb  & 1.6 fb
\\
\hline $\sigma^{\textrm{max}}_{J/\psi \mathcal{G}_0}/%
        \sigma_{J/\psi h }$
& 0.63    & 0.72     & 1.9
\end{tabular}
\end{ruledtabular}
\end{table}

\section{Discussion}
The anomalously large cross section for exclusive 
$J/\psi+\eta_c$ production measured by the BELLE Collaboration 
is not well understood within pQCD factorization or within the 
NRQCD factorization formalism.
If the BELLE signals are made purely of $J/\psi+\eta_c$ events, then 
either pQCD factorization fails or there are important production 
mechanisms that we have not yet taken into account. The former 
possibility would be a violation of an established pQCD factorization 
theorem.
If the exclusive two-charmonium production process is demonstrated to 
violate the factorization theorem, then one must understand why. Without 
a fundamental
understanding this problem, one can not safely use such factorization 
theorems, which are a central part of most current particle phenomenology.

\begin{acknowledgments}
We thank Geoffrey Bodwin, Eric Braaten, Stanley Brodsky, 
and Alfred Goldhaber for enjoyable collaborations in the work presented here.
JL is also grateful to Geoffrey Bodwin for many useful comments on
this manuscript. JL thanks KIAS for the invitation to the second ICFP
conference and hospitality during his stay. The research of JL in the
High Energy Physics Division at Argonne National Laboratory is supported
by the U.~S.~Department of Energy, Division of High Energy Physics, under
contract W-31-109-ENG-38.
\end{acknowledgments}


\end{document}